\documentstyle[prl,aps,floats,twocolumn]{revtex}
\begin{document}
\draft
\twocolumn[\hsize\textwidth\columnwidth\hsize\csname @twocolumnfalse\endcsname

\title{Phase diagram of the two-chain Hubbard model}
\author{
Youngho Park$^{1}$, Shoudan Liang$^{2}$\\
and T.K. Lee$^{1}$}
\address{${1})$ Institute of Physics, Academia Sinica, Nankang, Taipei 11529,
Taiwan\\
and National Center for Theoretical Sciences, P.O. Box 2-131, Hsinchu, Taiwan\\
${2})$ NASA Ames Research Center, Moffett Field, CA 94035, U.S.A.}

\date{\today}
\maketitle
\begin{abstract}

We have calculated the charge gap and spin gap for the two-chain Hubbard model
as a function of the on-site Coulomb interaction and the interchain 
hopping amplitude.
We used the density matrix renormalization group method and developed a method 
to calculate separately the gaps numerically for the symmetric 
and antisymmetric modes with respect to the exchange of the chain indices.
We have found very different behaviors for the weak and strong interaction 
cases.
Our calculated phase diagram is compared to the one obtained by Balents
and Fisher using the weak coupling renormalization group technique.

\end{abstract}
]

\newpage
\narrowtext

Although the Luttinger liquid behavior of the one-dimensional Hubbard model 
has been well understood, the two-dimensional Hubbard model,
which is believed to be related to understanding of the high T$_c$
superconductivity\cite{Anderson}, is not yet clear.
As a crossover between $1$D and $2$D, the two-chain Hubbard model is
certainly a good theoretical basis for the ladder compounds\cite{Dagotto,Scalapino}.
It is important to understand how two Luttinger liquid systems evolve
to the ladder system as the interchain coupling is introduced.

The Hamiltonian for the two-chain Hubbard model is
\begin{eqnarray}
H & = & -t_{||}\sum_{l,\langle i,j\rangle,\sigma}
     (c^\dagger_{li\sigma} c_{lj\sigma} + H.c.) 
    -t_{\perp}\sum_{i,\sigma}(c^\dagger_{1i\sigma} c_{2i\sigma} + H.c.)
\nonumber\\
& + & U\sum_{l,i} n_{li,\uparrow} n_{li,\downarrow},
\end{eqnarray}
where $l$ is the chain index, $l=1,2$.
We have the intrachain hopping $t_{||}$ term, the interchain hopping 
$t_{\perp}$ term and the on-site Coulomb interaction $U$ term.
$t_{||}=1$ in this paper.

Some authors have studied the phase diagram of the two-chain Hubbard
model using the weak coupling renormalization group method 
\cite{Fabrizio,Khveshchenko,Balents}.
At half-filling, the system is an insulator with a spin gap.
Upon light hole doping, for small $t_{\perp}$ the spin gap
remains finite.
For large $t_{\perp}$, the complete separation between the bonding band 
and the antibonding band leads the system to a Luttinger liquid phase.
But it remains unclear for large $U$ case which is equivalent to 
the $t-J$ ladder\cite{Dagotto2,Sigrist,Tsunetsugu,Poilblanc}.
Balents and Fisher described the phase
in terms of number of gapless charge and spin modes, which they denoted by
CnSm where n is the number of gapless charge modes and m is the number of 
gapless spin modes\cite{Balents}.
For our work we also use this notation.
Their phase diagram is rather diverse in the hole-doped region.
Interestingly, they found phases such as C$2$S$2$ and C$2$S$1$ 
between the C$1$S$1$ Luttinger liquid phase and the spin-gapped C$1$S$0$ phase.
And the phase diagram is qualitatively same for different electron filling 
$n$ except for the cases of half-filling, quater-filling and half-filled 
bonding band, where the umklapp process will be relevant.

Since most of the previous results done by the weak coupling renormalization
group method may be reliable only in the $U\rightarrow 0^+$ limit, the
questions are whether it will be similar for finite $U$ and
whether it will depend on $U$.
Noack {\it et al.} have studied the correlation functions and the gaps
for finite $U$ using the density matrix renormalization group method
(DMRG)\cite{Noack}.
They found the enhancement of the d-wave pairing correlation function for
the spin-gapped phase at rather anisotropic regime $t_{\perp}>t_{||}$, 
though they cannot distinguish the symmetric and antisymmetric 
modes.
Hence it is difficult to determine which CnSm phase it is.
To understand the phase diagram in detail and to make comparison with 
the weak coupling result it is necessary to study the charge and spin 
excitations in different modes for finite $U$.

In this paper, we study the phase diagram of the two-chain Hubbard model
by calculations of the charge and spin gaps for the symmetric mode and 
antisymmetric mode respectively for finite values of $U$
at a fixed electron filling $n=0.75$.
Using this result we can easily understand the result for other electron
filling.
For our study, we use the density matrix renormalization group method 
(DMRG)\cite{White,Liang}
and develop a method to differentiate the modes.

We take the transform of $c$ operators in terms of bonding and
antibonding forms as following:
\begin{equation}
c_{\pm i\sigma} = \frac{1}{\sqrt{2}}(c_{1i\sigma} \pm c_{2i\sigma}).
\end{equation}
\newline
Then the Hamiltonian becomes
\begin{eqnarray}
H & = & -t_{||} \sum_{\langle i,j\rangle,\sigma}
        (c^\dagger_{+i\sigma} c_{+j\sigma} + 
         c^\dagger_{-i\sigma} c_{-j\sigma} + H.c.) \nonumber\\
  &   & -t_{\perp} \sum_{i,\sigma} (c^\dagger_{+i\sigma} c_{+i\sigma} 
                            -c^\dagger_{-i\sigma} c_{-i\sigma}) \nonumber\\
  &   & +\frac{U}{2}\sum_{i} [ (n_{+i\uparrow}+n_{-i\uparrow})
                               (n_{+i\downarrow}+n_{-i\downarrow}) \nonumber\\
  &   & + c_{+i\uparrow}^\dagger c_{+i\downarrow}^\dagger 
          c_{-i\downarrow} c_{-i\uparrow}
        + c_{-i\uparrow}^\dagger c_{-i\downarrow}^\dagger 
          c_{+i\downarrow} c_{+i\uparrow} \nonumber\\
  &   & + c_{+i\uparrow}^\dagger c_{-i\downarrow}^\dagger 
          c_{+i\downarrow} c_{-i\uparrow}
        + c_{-i\uparrow}^\dagger c_{+i\downarrow}^\dagger 
          c_{-i\downarrow} c_{+i\uparrow}].  
\end{eqnarray}
Besides the usual symmetry of the total spin $S$ and $S_z$, the translational
symmetry along the chain direction, the Hamiltonian also has the symmetry
of the exchange of the chain indices.
The symmetry of the total state under the exchange of two chain indices
depends on the symmetries of wavefunction of each particle.
It is decided by the number of particles on the odd chain of the new 
transformed Hamiltonian, since the evenness and oddness of the number of 
particles on the odd chain is not changed by the Hamiltonian.
If the number of particles on the odd chain is odd, total state is 
antisymmetric and if the number is even, the total state is symmetric.  
In the DMRG calculation, we define the symmetry under the chain exchange of 
each base or sector in terms of the number of particles on the odd chain and 
so we 
have the option to choose the symmetry and able to calculate the ground state 
energy of a given symmetry.
We also use the open boundary condition here hence there is no translational
symmetry along the chain direction.

For a set of given values of $U$, $t_{\perp}$ and electron 
filling $n$, the charge gaps and spin gaps are defined as follows:
\begin{eqnarray}
&\Delta_{c+} = \frac{1}{2}[E_{+}(Q+2, S=S_z=0)\nonumber\\
&+E_{+}(Q-2, S=S_z=0)-2 E_{+}(Q, S=S_z=0)],
\end{eqnarray}
\begin{equation}
\Delta_{c-} = E_{-}(Q, S=S_z=0)-E_{+}(Q, S=S_z=0),
\end{equation}
\begin{equation}
\Delta_{s+} = E_{+}(Q, S=S_z=1)-E_{+}(Q, S=S_z=0),
\end{equation}
\begin{equation}
\Delta_{s-} = E_{-}(Q, S=S_z=1)-E_{+}(Q, S=S_z=0),
\end{equation}
where $\pm$ is symmetric or antisymmetric mode and all energies are the
lowest energy for each set of quantum numbers.  
Here, we assume the total number of the particles is even.
In each iteration of the DMRG calculation, we typically keep $M=200$ states
for the block.  We calculated gaps for various
size of lattices with length $L=8$, $L=12$, $L=16$ and $L=20$ and 
extrapolated the thermodynamic values by $1/L$ polynomial expansions.
In the DMRG calculation, we only utilize the conservation of
$S_z$, not the total spin $S$.
To identify the total spin $S$ correctly, it is necessary to check $S$ values 
of the lowest energy state in each sector.
For $S_z=0$ sector with symmetric mode, the ground state is $S=0$ state.
For $S_z=1$ sector with symmetric mode, the ground state is $S=1$ state and
the same is true for $S_z=1$ sector with antisymmetric mode.
But for $S_z=0$ sector with antisymmetric mode, the ground state is $S=1$
state rather than $S=0$, which implies $\Delta_{c-}\geq\Delta_{s-}$ for
finite size lattices.
Therefore, to calculate 
the antisymmetric charge gap we need to find the $S=0$ state which is an
excited state.  We find this state by calculating the 
expectation values of $S$ for a few excited states.

\vskip -1.0cm
\begin{figure}[h]
\includegraphics{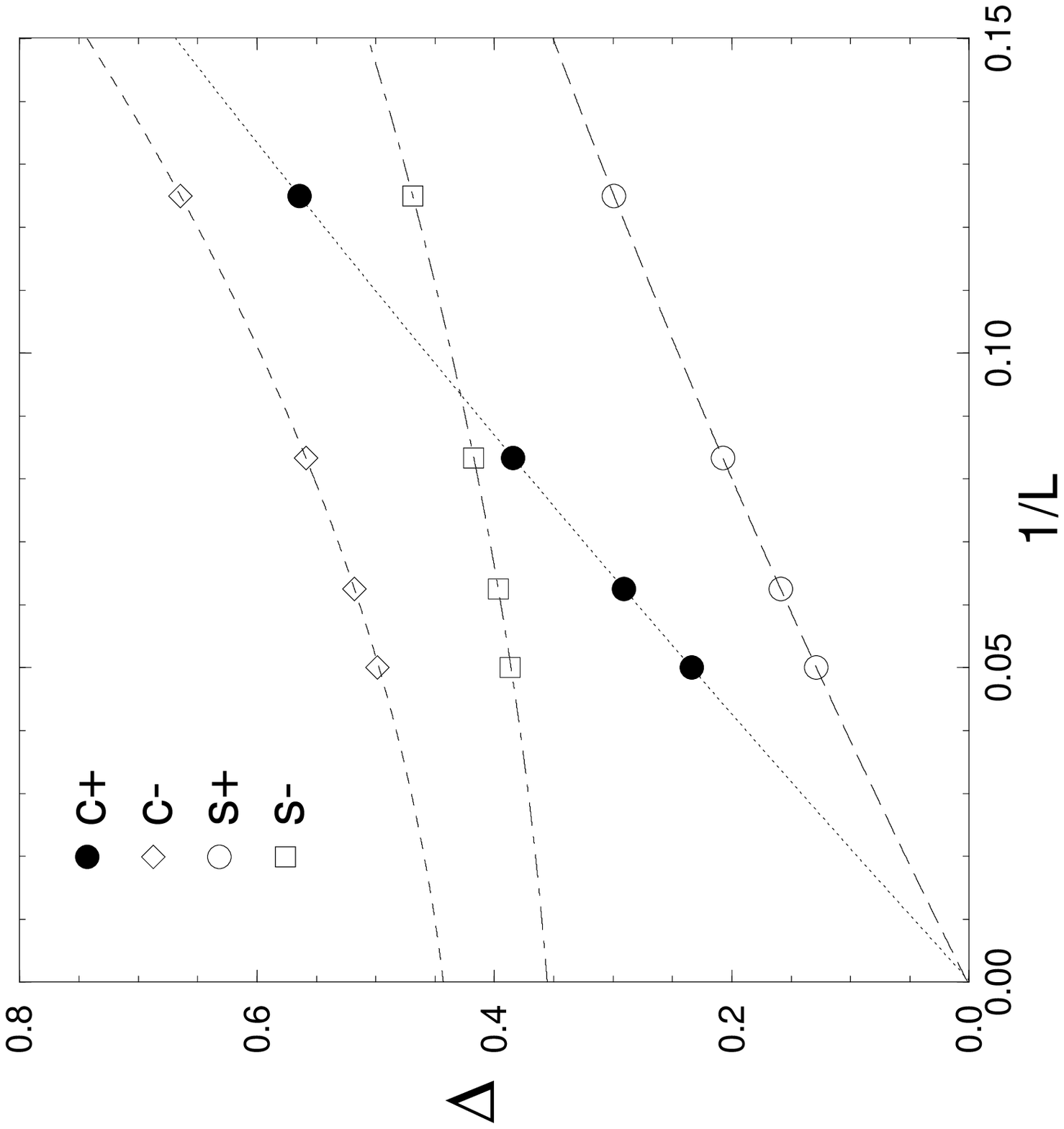}
\end{figure}
\vskip 8.8cm

\noindent{FIG. 1. 
Gaps versus $1/L$ to calculate the thermodynamic values of gaps 
for $U=8$ and $t_{\perp}=1.7$.
We denote symmetric charge gaps with dark circles, antisymmetric
charge gaps with diamonds, symmetric spin gaps with open circles
and antisymmetric spin gaps with squares.
}
\vskip 0.7cm

Fig. $1$ shows the finite size extrapolation used to obtain the 
thermodynamic values of the gaps for $U=8$ and $t_{\perp}=1.7$. 
The symmetric charge gap $\Delta_{c+}$ and the symmetric spin gap 
$\Delta_{s+}$ vanish.  Specially, for the entire range of $U$ and $t_{\perp}$,
$\Delta_{c+}$ always vanishes at $n=0.75$.
The antisymmetric charge gap $\Delta_{c-}$ and the antisymmetric spin gap 
$\Delta_{s-}$ are finite.
$\Delta_{c-}$ is larger than $\Delta_{s-}$ for all size of lattices
but, for the longer lattice the difference between them is smaller.
Since we take the ground state as the target state for the calculations
of the density matrix in DMRG, the accuracy of the excited states is
not as high as that of the ground state and this is even more serious for
longer lattice and for small $t_{\perp}$.

For strong interaction $U=8$, $\Delta_{s+}$ and $\Delta_{s-}$ are 
plotted in Fig. $2$. as functions of $t_{\perp}$ for various size of lattices 
including $L=\infty$ extrapolated values.
For most values of $t_{\perp}$ we get very good extrapolations for both 
$\Delta_{s+}$ and $\Delta_{s-}$with three $L$ values: $8$, $12$ and $16$.
There are some points near $t_{\perp}=1.0$ for which we are not able to get 
a good extrapolation.
As $t_{\perp}$ decreases from $t_{\perp}=2.0$ $\Delta_{s-}$ decreases
but does not vanish until $t_{\perp}=0.6$, and $\Delta_{s+}$ opens up
around at $t_{\perp}=1.5$.  
Since we always have one gapless symmetric charge mode, that is 
$\Delta_{c+}=0$ and $\Delta_{c-}\geq\Delta_{s-}$ the phase changes 
from C$1$S$1$ phase to C$1$S$0$ phase around $t_{\perp}=1.5$.   
Between these two phase we do not have phases such as C$2$S$2$ and
C$2$S$1$ which are found in the results for $U\rightarrow 0^+$ limit 
by Balents and Fisher \cite{Balents}.
We stop our calculations at $t_{\perp}=0.6$ because the DMRG calculation
has poor accuracy for small $t_{\perp}$.

\vskip -0.8cm
\begin{figure}[h]
\includegraphics{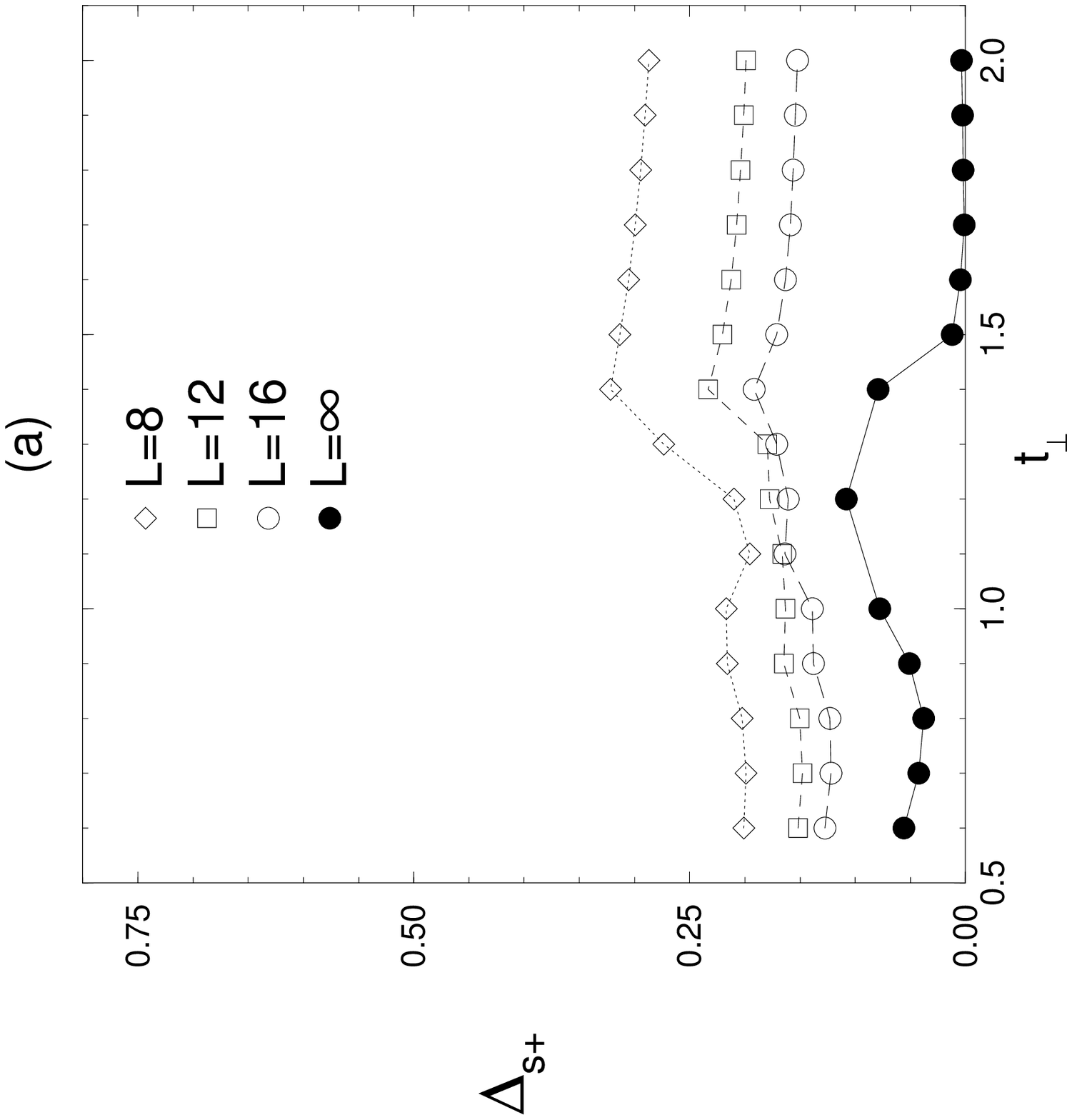}
\end{figure}
\vskip 8.2cm

\vskip -1.0cm
\begin{figure}[h]
\includegraphics{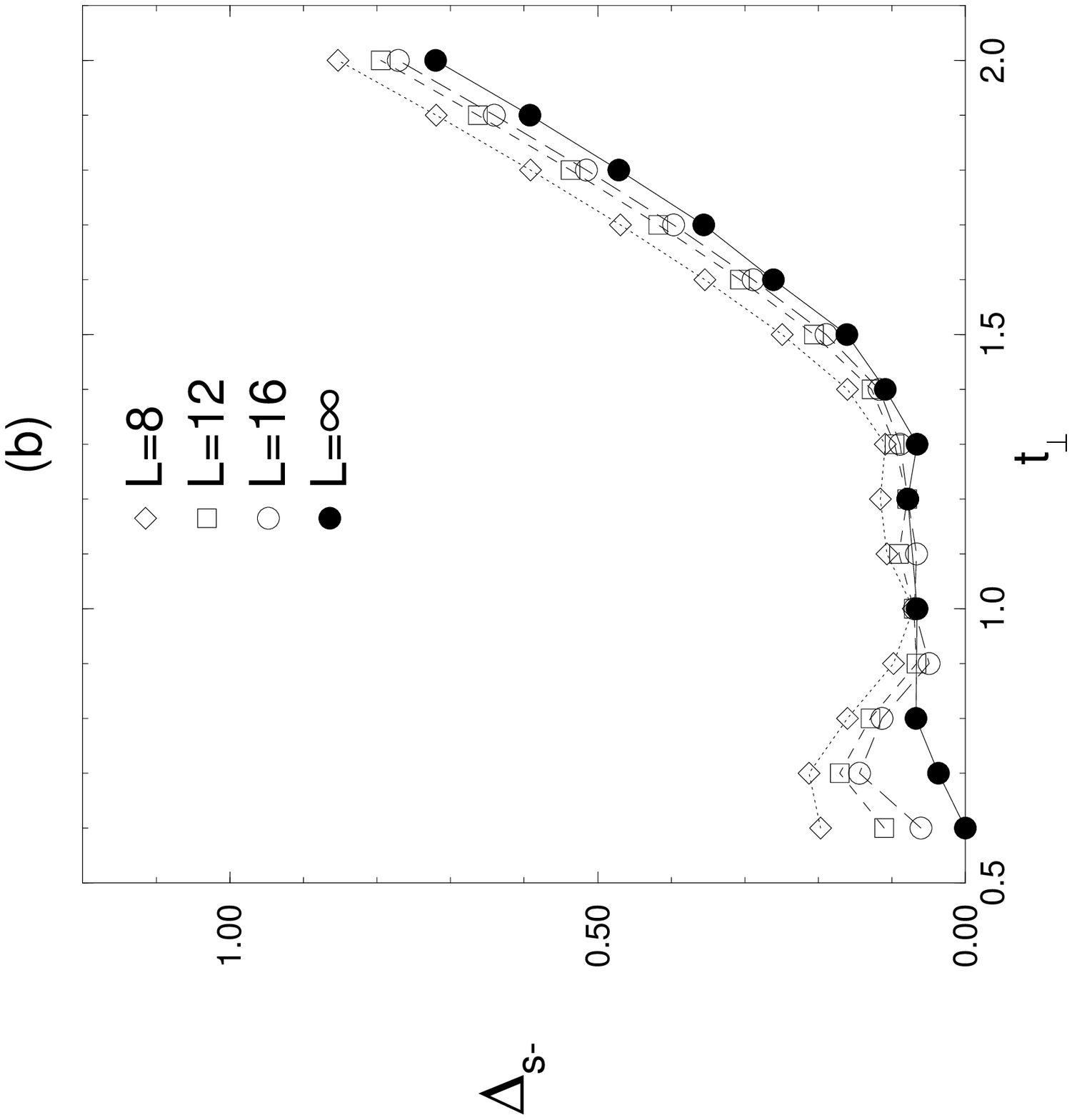}
\end{figure}
\vskip 8.8cm

\noindent{FIG. 2. 
Spin gaps versus  $t_{\perp}$ for $U=8$.
(a) for symmetric mode and (b) for antisymmetric mode.
We denote values for $L=8$ with diamonds, $L=12$ with squares,
$L=16$ with open circles and $L=\infty$ extrapolated values with
dark circles.
}
\vskip 0.6cm

The vanishing of $\Delta_{s-}$ at $t_{\perp}=0.6$ can be understood from 
the $t_{\perp}=0$ limit. In this limit, two chains will be completely
decoupled to two one-dimensional chains, which is the Luttinger phase, 
and we will have C$2$S$2$ phase with all gapless modes.
So, as $t_{\perp}$ decreases all gaps will vanish one by one and eventually
we will have C$2$S$2$ phase in the $t_{\perp}=0$ limit.
The linear behavior of $\Delta_{s-}$ in the C$1$S$1$ phase region for 
$t_{\perp}>1.5$ can be understood in the limit of infinitely large $t_{\perp}$.
In this limit, $\Delta_{s-}$
measures the separation between bonding band and empty antibonding band.
In the physically interesting isotropic region $t_{\perp}=1.0$, both spin
gaps are similar in magnitude but finite, which is consistent with previous 
results\cite{Noack}.
Also, they increase slowly as $t_{\perp}$ increases.
This is compatible with the results for the $t-J$ ladder, which is linear
behavior of the spin gap in $J_{\perp}$ ( $\Delta_s\sim J_{\perp}\sim
t_{\perp}^2/U$ )\cite{Dagotto2,Tsunetsugu}.
The increasing behavior of $\Delta_{s+}$ for $t_{\perp}<0.7$ is an artifact
from poor convergence of DMRG for small $t_{\perp}$.

\vskip -0.8cm
\begin{figure}[h]
\includegraphics{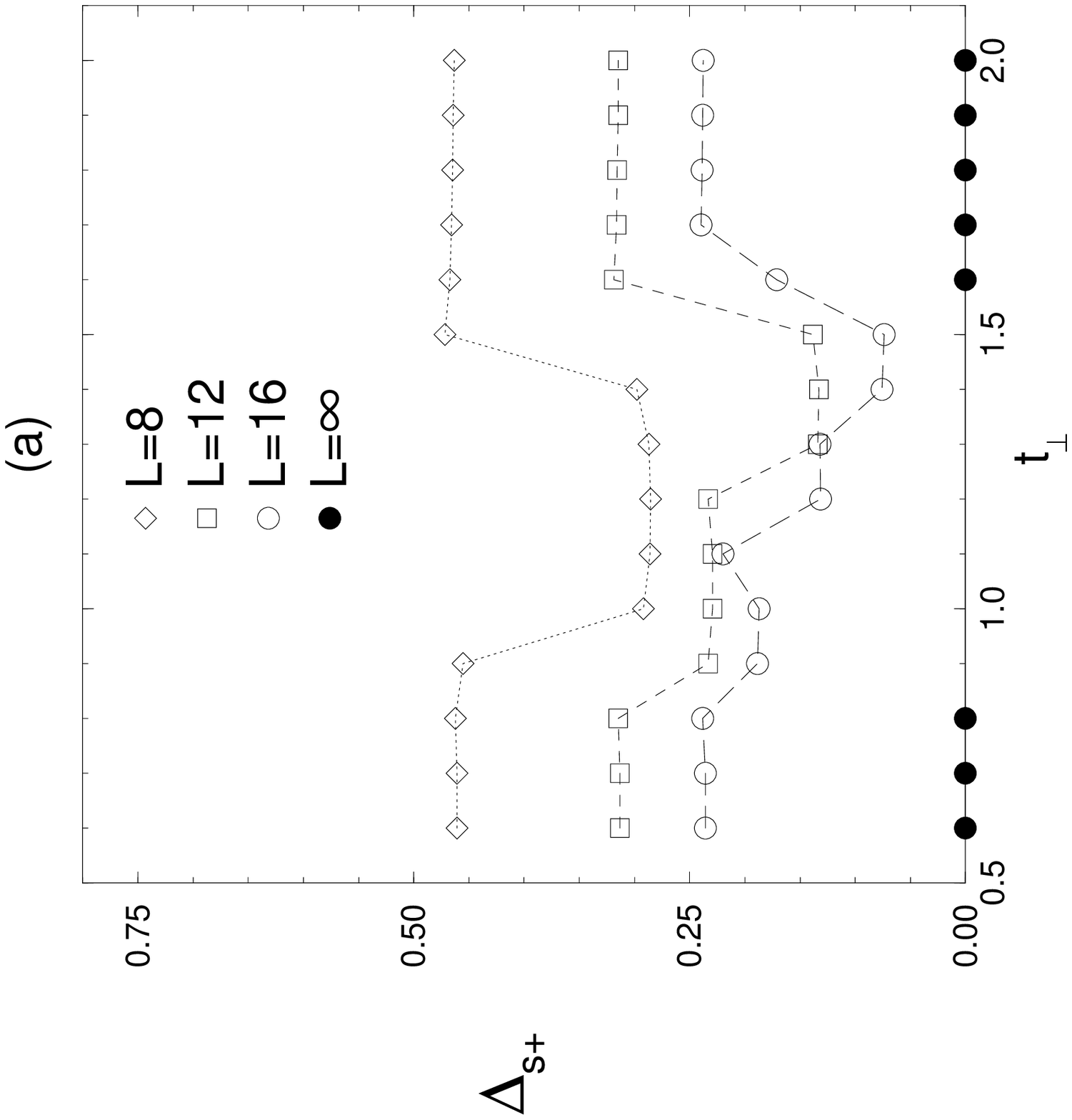}
\end{figure}
\vskip 8.2cm

\vskip -1.0cm
\begin{figure}[h]
\includegraphics{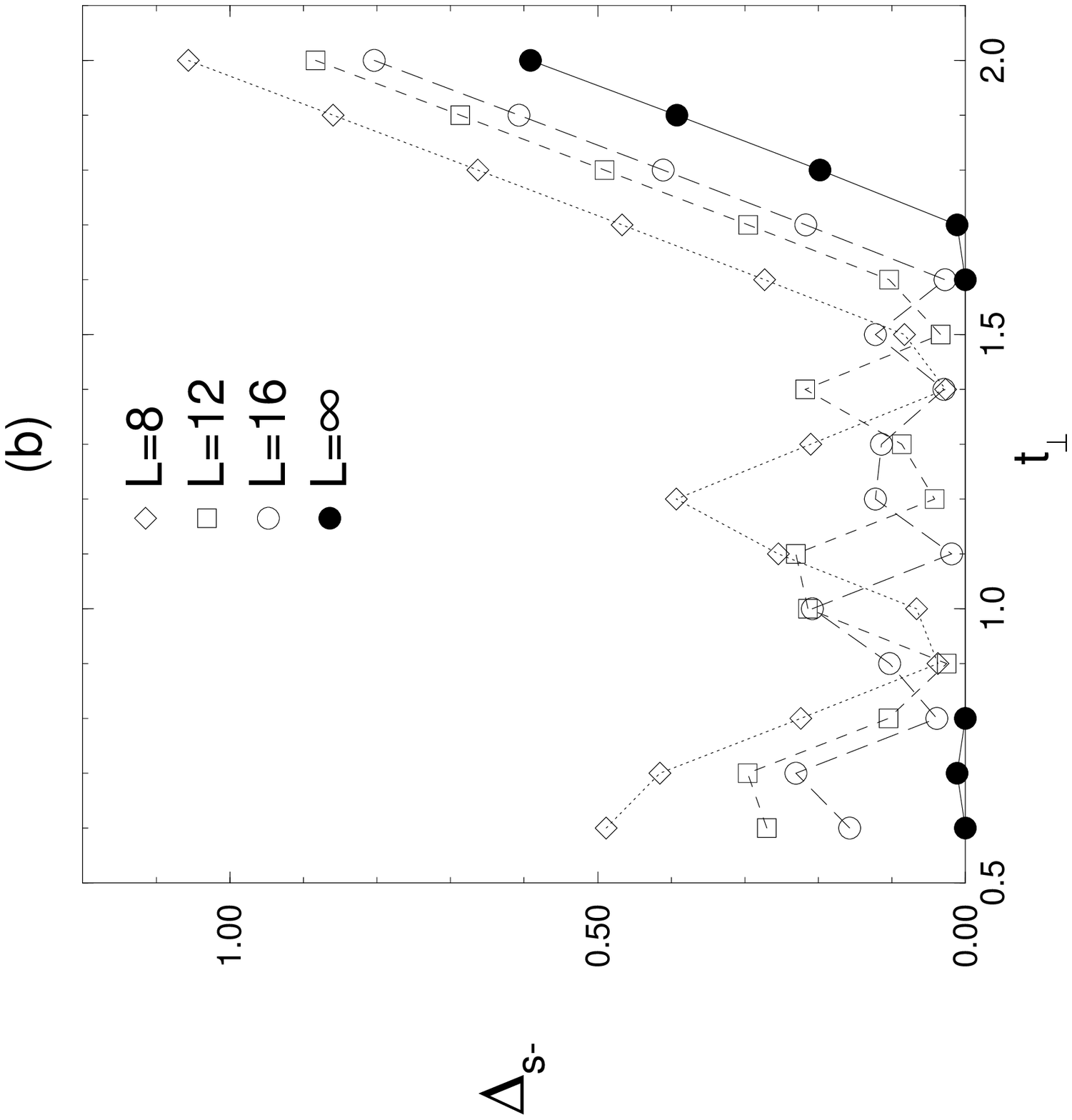}
\end{figure}
\vskip 8.8cm

\noindent{FIG. 3. 
Same figures as FIG.2 for $U=1$.
}
\vskip 0.6cm

Fig. $3$ shows same figures for weak interaction $U=1$ as Fig. $2$.
Unlike $U=8$ case, $\Delta_{s+}$ does not open up and $\Delta_{s-}$ 
vanishes around at $t_{\perp}=1.7$, where $\Delta_{c-}$ also vanishes
so that we have a transition from C$1$S$1$ phase to C$2$S$2$ phase.
As $t_{\perp}$ decreases further, we have finite size fluctuations
so that we can not calculate $L=\infty$ gaps until $t_{\perp}=0.8$.
Surprisingly, in a small region of $t_{\perp}$, $0.6\leq t_{\perp}\leq 0.8$,
there is little finite size fluctuation.
We find both $\Delta_{s+}$ and $\Delta_{s-}$ vanish.
On the other hand $\Delta_{c-}$ is finite, but 
considering the overestimation of $\Delta_{c-}$ because of the inaccuracy 
of the excited states and the system being two decoupled Luttinger liquid
systems for small $t_{\perp}$, presumably it also vanishes and we have
C$2$S$2$ phase.
If we smoothly connect this region to the region with $t_{\perp}\simeq 1.7$,
it is plausible to assume that the phase remains gapless.
Therefore, in this weak interaction case, there will be only one transition
from C$1$S$1$ phase at large $t_{\perp}$ to C$2$S$2$ phase at small 
$t_{\perp}$ and this is rather consistent with the noninteracting limit.
In the C$1$S$1$ phase, we also have the linear behavior of $\Delta_{s-}$
with smaller $\Delta_{s-}$ than $U=8$ case but with larger slope.

\vskip -1.0cm
\begin{figure}[h]
\includegraphics{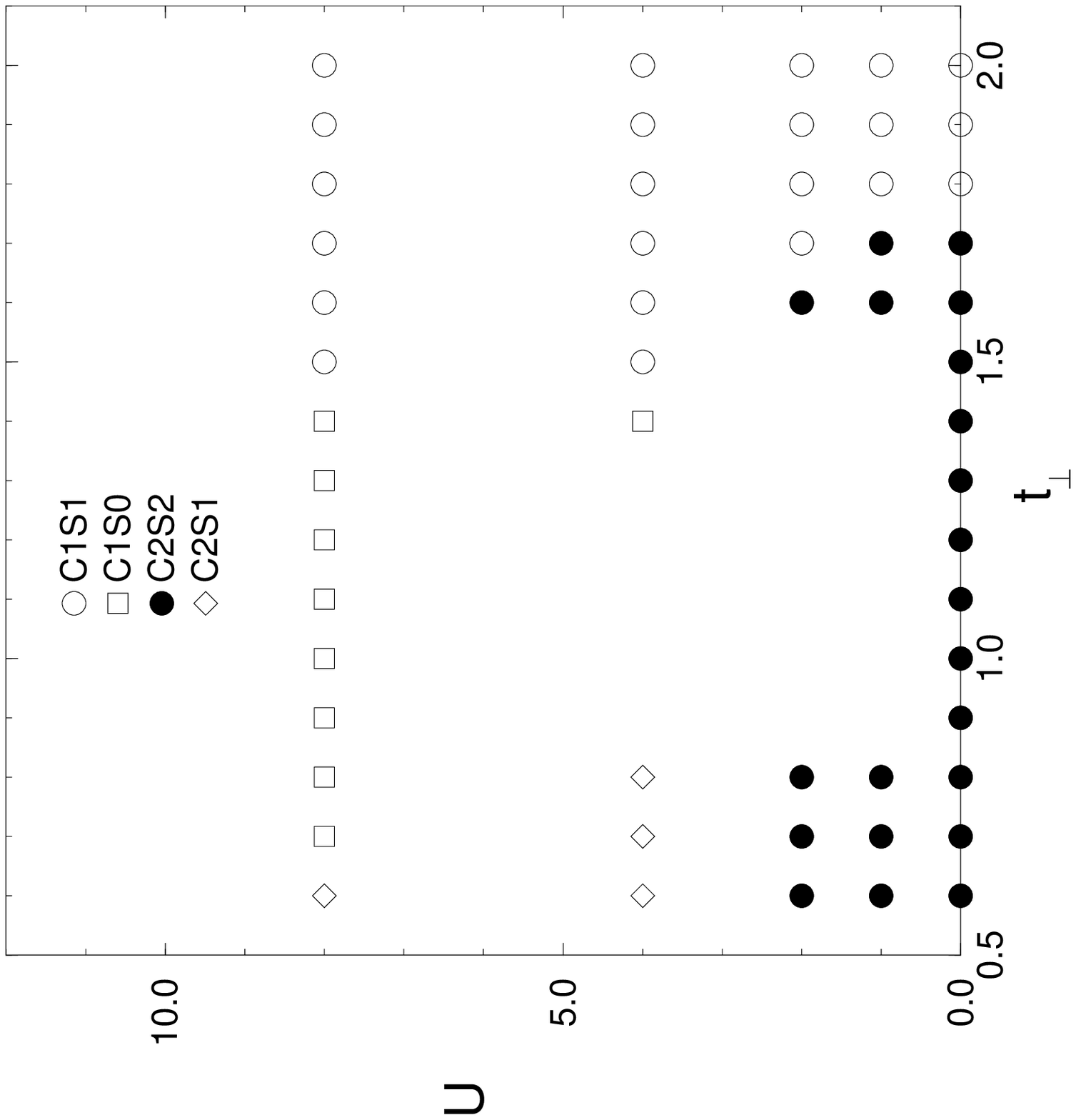}
\end{figure}
\vskip 8.8cm

\noindent{FIG. 4. 
Phase diagram at $n=0.75$.
We denote C$1$S$1$ phase with open circles, C$1$S$0$ with squares,
C$2$S$2$ with dark circles and C$2$S$1$ with diamonds.
We also show the phases for the noninteracting
case.
}
\vskip 0.7cm

With calculations for other values of $U$, we plot the phase diagram
in the space of $t_{\perp}$ and $U$ at $n=0.75$ in Fig. $4$.
It is clear that the strong interaction case ($U=8$) and the weak interaction
case ($U=1$) are clearly different.
$U=2$ case is same as $U=1$ case and $U=4$ seems to be similar to $U=8$ case
regardless of the existence of the finite size fluctuations.
The transition point of $t_{\perp}$ between two different phases decreases
from $t_{\perp}\simeq 1.7$ for the noninteracting case as $U$ increases.
When we have different $n$, the transition point of $t_{\perp}$ will be 
changed because it is around the value of the noninteracting limit,
$t_{\perp}=1-\cos \pi n$ just as the weak coupling case\cite{Balents}.
Therefore, for different $n$ the whole phase diagram will be shifted in
$t_{\perp}$ direction.

In conclusion, we have derived the phase diagram of the two-chain
Hubbard model in terms of charge and spin excitations for both symmetric 
and antisymmetric modes at fixed electron filling.
We have found very different behaviors for the weak and strong interaction 
cases.
The transition from the noninteracting limit to the interaction case seems to
be rather gradual and beyond finite value of $U$ we have the spin-gapped
phase at the isotropic region $t_{\perp}=1.0$, which is consistent with
the previous results for the strong interaction case and $t-J$ model.

\indent

This work is partially supported by the Office of Naval Research
Contract Nos. N00014-92-J-1340 and N00014-95-1-0398
and by the National Science Foundation Grant No. DMR-9403201,
and by the National Science Council, Rep. of China, Grant Nos.
NSC87-2112-M-011-016.

\end{document}